\documentclass[10pt, conference, compsocconf,letter]{IEEEtran}
\usepackage{graphicx}
\usepackage{amsfonts}
\usepackage{amssymb,amsmath}
\usepackage{euscript}
\usepackage{color}
\usepackage{slashbox}
\usepackage{url}
\usepackage[marginpar]{todo} 

\DeclareMathOperator{\hash}{\mathrm{Hash}}

\begin{document}
\title{State Space Exploration of RT Systems in the Cloud}
\author{\IEEEauthorblockN{Carlo Bellettini, Matteo Camilli, Lorenzo Capra, Mattia Monga}
\IEEEauthorblockA{Dipartimento di Informatica e Comunicazione\\
Universit\`a degli Studi di Milano\\
Milano, Italy\\ 
Email: \{bellettini, camilli, capra, monga\}@dico.unimi.it}}
\maketitle

\begin{abstract}
The growing availability of distributed and cloud computing  frameworks make it possible to face complex computational problems
in a more effective and convenient way.
A notable example is state-space exploration of discrete-event systems specified in a formal way.

The exponential complexity of this task is a major limitation to the usage of consolidated
analysis techniques and tools.
We present and compare two different approaches to state-space explosion,
relying on distributed and cloud frameworks, respectively. 
These approaches were designed and implemented following the same computational schema, a sort of map \& fold.
They are applied on symbolic state-space exploration of real-time systems specified by (a timed extension of) 
Petri Nets, by readapting a sequential algorithm implemented as a command-line
\texttt{Java} tool.
The outcome of several tests performed on a benchmarking specification are presented, thus
showing the convenience of cloud approaches.


\end{abstract}

\section{Introduction}
\label{sec:intro}

State-space exploration is the most widely used technique for the analysis of discrete-event systems specified in a formal way,
due to the completeness of provided information, and the possibility of being easily automated.
However, a known major weakness of this
approach is the possible combinatorial growing of state-space with respect to models' size~\cite{Valmari98}.

A typical application area of state-space exploration is the validation of Real-Time (RT) systems, that
require intensive verification before deployment.
Several formal models for RT systems have been proposed~\cite{Furia10}, among which time extensions of Petri nets (PN) play an important role. The verification of RT properties, that mix logical 
%
and timing
aspects, usually requires building directed graphs expressing the system behavior in terms of state-transitions~\cite{Bellettini11,Furia10}, starting from an initial state.
RT constraints make this an even more challenging task. In the case of a dense time domain, 
the set of reachable states is likely to be infinite: this is normally
tackled by clustering classes of states which share some reachability and timing conditions~\cite{Bellettini11,Furia10}.
Yet, time breaks the locality of events' occurrences that is a key feature in
classical state-space exploration techniques.

We introduce and compare two different approaches to state-space exploration,
based on distributed and cloud computing frameworks.
Although these approaches do not alleviate state-space explosion, they lead to a significant speed-up of execution times (by considerably increasing the storage space and computation power at disposal) and permit computing resources to be scaled up.

In accordance to a consolidated idea, independent processing units (sw or hw) 
are in charge of building partitions of the state-transition  graph, synchronizing at the end of the computation
in order to consistently compose the whole structure.
What characterizes our approaches, making them parametric to the adopted formalisms,
is a full adherence to a computational pattern which lies in
iterating a sequence of elementary ``map-fold'' operations. For example they could be easily specialized to work with
different kinds of PNs, or they could be exploited in the context of model-checking for efficiently translating Labeled Transition System (LTs) from an implicit representation to an explicit one \cite{Garavel01}.

Our reference model is Time-Basic (TB) nets~\cite{UnifiedWay91}, an expressive formalism for RT systems' specification.
An efficient state-space exploration technique for TB nets was recently  implemented as a sequential Java program~\cite{Bellettini11}.
The output is a symbolic
state-transitions graph (TRG),
that overcomes
the old analyzer of TB nets~\cite{IPTES-PDM41} (based in turn on a time-bounded inspection of a symbolic \emph{tree}).
In this paper we present how we have adapted the sequential TRG builder in
order to exploit distributed/cloud computing frameworks.  
A summary of test sessions carried out on a benchmarking system specification (the Gas Burner~\cite{Ravn93,IPTES-PDM41}) is also included. The proposed approaches are shown to significantly improve the sequential algorithm performances,
both in terms of execution time and analyzable model's size.


\subsection{TB nets and timed reachability analysis}
\label{sec:analysis}

TB nets~\cite{UnifiedWay91} belong to the category of formalisms
in which time constraints on systems' state transitions 
are expressed as numerical intervals,
denoting the possible instants at which some events may occur.
Intervals' domain is assumed here $\mathbb{R}^+$.
TB nets are very expressive, for two main reasons:
first, interval bounds are functions of the time description of a state; 
secondly, each event occurrence may be assigned either a \emph{weak}  or a \emph{strong} semantics:
under some conditions, a given event either \emph{may} or \emph{must} occur.

Let us recall a few computationally relevant points of the TRG
algorithm proposed in~\cite{Bellettini11}. We here omit unessential details related to the employed formalism.

A TRG node represents a symbolic state $S=\langle M,C\rangle$, where 
$M$ is the topological  description of a system state 
given in terms of symbols denoting time-stamps (a \emph{marking,}
following the PNs parlance\footnote{More precisely, $M$ is defined by a finite set of \emph{places}, each associated
to a multi-set of symbols, called \emph{tokens.}}),
$C$ is a predicate expressed as linear inequalities involving such symbols.
Assuming no absolute time references are used in a TB model, $C$ only contains \emph{relative} time dependencies, e.g., $T_2 - T_1 \leq 1.5 \wedge T_0 \leq T_1$.
The most expensive task is verifying inclusion between nodes, meant as classes of corresponding ordinary states:
when a successor $S'$ of node $S$ is generated, we check whether any node $S''$ already exists such that either $S' \subseteq S''$
or  $S' \supset S''$ (in the latter case $S'$ absorbs $S''$,  and it is set ``to be processed'').  
A symbolic state normalization is required, involving different actions.
First, symbols occurring in $C$, but not in $M$, are
eliminated. What constitutes the key point of the whole
algorithm~---~and very often enables termination~---~however, is a
quite sophisticated procedure able to recognize symbols
that are irrelevant for the model evolution. Such symbols are 
replaced in $M$ by anonymous time-stamps, then they are possibly eliminated from $C$.
Let $S$, $S'$ be normalized:
a sufficient condition for $S \subseteq S'$ is $M = M'$ and $C \wedge \neg C' \equiv false$ \footnote{The Floyd-Warshall and the Simplex algorithms are used
for variable elimination and satisfiability check, respectively.}.

\section{Sequential model}
\label{sec:seq}

\begin{figure}[htbf]
\centering
\includegraphics[width=0.8\columnwidth]{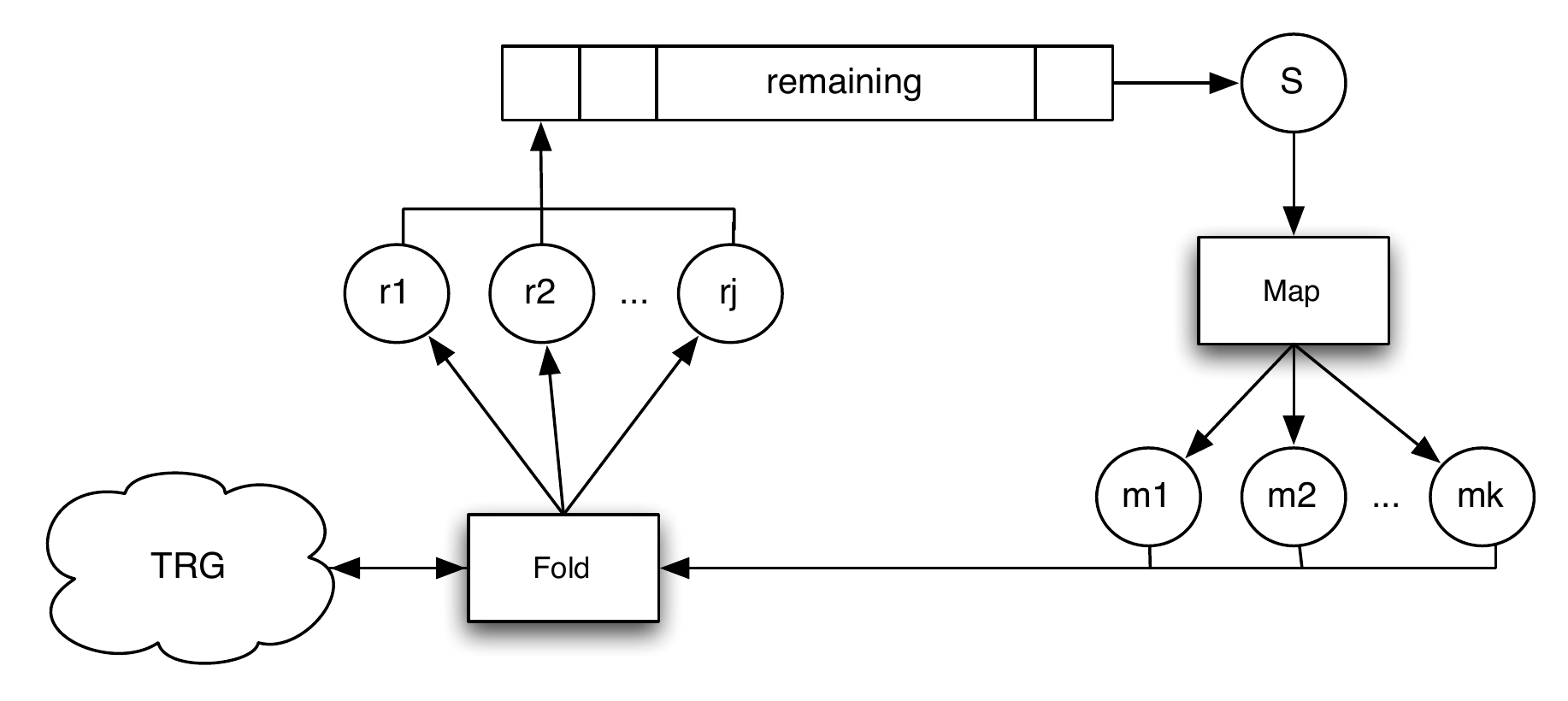}
\caption{Sequential model}
\label{fig:seqmodel}
\end{figure}

The TRG construction has been automated by means of a Java tool called $Graphgen$. The corresponding sequential algorithm is sketched in the Fig.~\ref{fig:seqmodel}.

The \texttt{remaining} list contains the reachable nodes of the graph
not yet examined, i.e., the \emph{expansion front} of the graph.  The graph builder takes one node at a time from the expansion front and executes two main phases: \texttt{Map} and \texttt{Fold}.
These operations derive from a well known programming model in which a
Map instance takes as input a sequence of values and computes a given
function for each value. Then, a Fold instance combines in some way
the elements of the sequence using an associative binary operation. 

In the TRG builder, the \texttt{Map} generates the successors of a node, the \texttt{Fold} combines them with the already existing nodes by identifying possible
inclusion relationships. Whenever the \texttt{Fold} phase identifies a relation between a new node $A$ (just computed by the \texttt{Map}) and an old node $B$ (already expanded), different operations must be performed on the adjacent edges depending on the relation between $A$ and $B$.

\begin{itemize} 
\item If $A \subseteq B$, the incoming edges of $A$ are redirected to $B$. The outgoing edges are not yet calculated, thus no actions are required.
\item If $A \supset B$, the incoming edges of $B$ are redirected to $A$ and the outgoing edges of $B$ (subset of the A's ones) are removed. 
\end{itemize}

At the end of the \texttt{Fold} phase the nodes computed by the \texttt{Map} which are not included in any old nodes, are placed into the \texttt{remaining} list. The \texttt{Map} phase and the \texttt{Fold} phase are repeated until the expansion front becomes empty.




\section{Parallel models}
\label{sec:parallelmod}

The sequential TRG builder execution takes more than 7 hours even for
a relatively small example as the Gas Burner is.
It is however possible to identify independent computational
sequences, in order to exploit the TRG algorithm in multi-thread and
distributed frameworks. We conceived two different ways for organizing
parallel computations. The two models are described in the following.

\paragraph{Workers model}
\label{sec:first-model}

\begin{figure}[htbf]
\centering
\includegraphics[width=0.9\columnwidth]{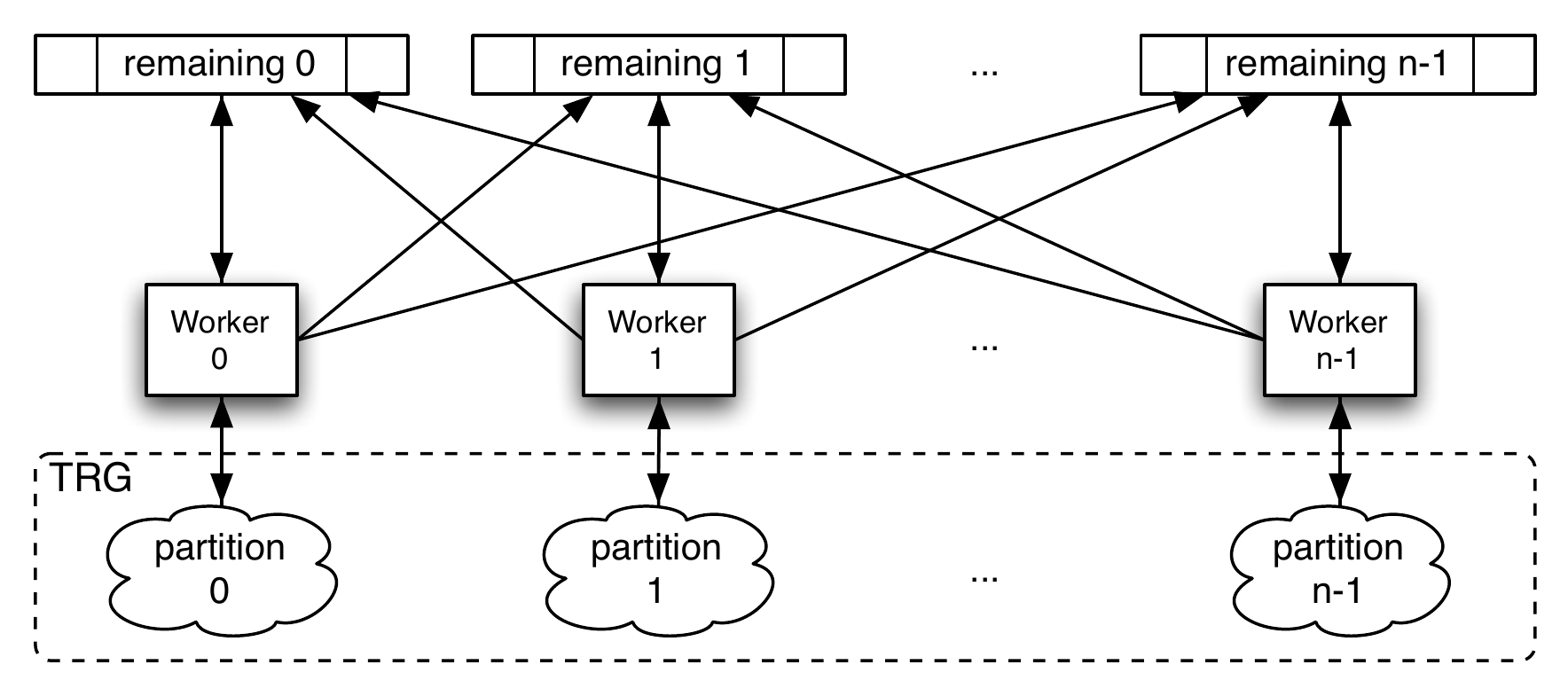}
\caption{\texttt{Worker}s model}
\label{fig:mt1}
\end{figure}

This model parallelizes the processing of nodes in the
expansion front. A set of independent computational units
(\texttt{Worker}s, see Fig.~\ref{fig:mt1}) locally execute the
\texttt{Map} and \texttt{Fold} phases. Each \texttt{Worker} computes a
portion of the final graph by examining a set of \emph{similar} nodes.
The whole state space is partitioned among the \texttt{Worker}s by
applying to each reachable state $S$ the following function:
 
\begin{equation} \label{eq:partitioning}
\hash(f(S)) \bmod n
\end{equation}
where $n$ is the number of \texttt{Worker}s, and $f$ extracts some
features from $S$ ensuring that the equality of $f(S)$ is a necessary
condition for inclusion relationships. In our implementations, $f$ is an easy to compute abstraction on $M$~---~called \emph{soft marking}~---~such that
the equality of soft markings is a necessary condition for two symbolic states to be included
into one another. The first definition of soft marking we used disregards the identity of time-stamp symbols.
Let $|M(p)|$ be the number of tokens in the place $p$. The soft marking of a state $S $ 
is defined as:

\begin{equation} \label{eq:soft_marking}
\begin{split}
f(S) = \langle |M(p_{1})|, ..., |M(p_{k})|  \rangle \,\in \mathbb{N}^{k}
\end{split}
\end{equation}

where $p_{1},...,p_{k}$ are the places of the analyzed TB net. 


%
%

Thus, any two nodes possibly related by inclusion are assigned to the same \texttt{Worker}.
Therefore, each \texttt{Worker} is able to locally accomplish the fold operation. Then it
sends the mapped nodes for which it is not responsible to the
appropriate peers. Fig.~\ref{fig:mt1} shows the overall architecture of
this model: each \texttt{Worker} has its own \texttt{remaining} list,
which contains nodes not yet examined. The expansion front is now the
overall union of all local \texttt{remaining} lists.

\paragraph{Mappers \& Folders model}
\label{sec:second-model}

\begin{figure}[htbf]
\centering
\includegraphics[width=0.9\columnwidth]{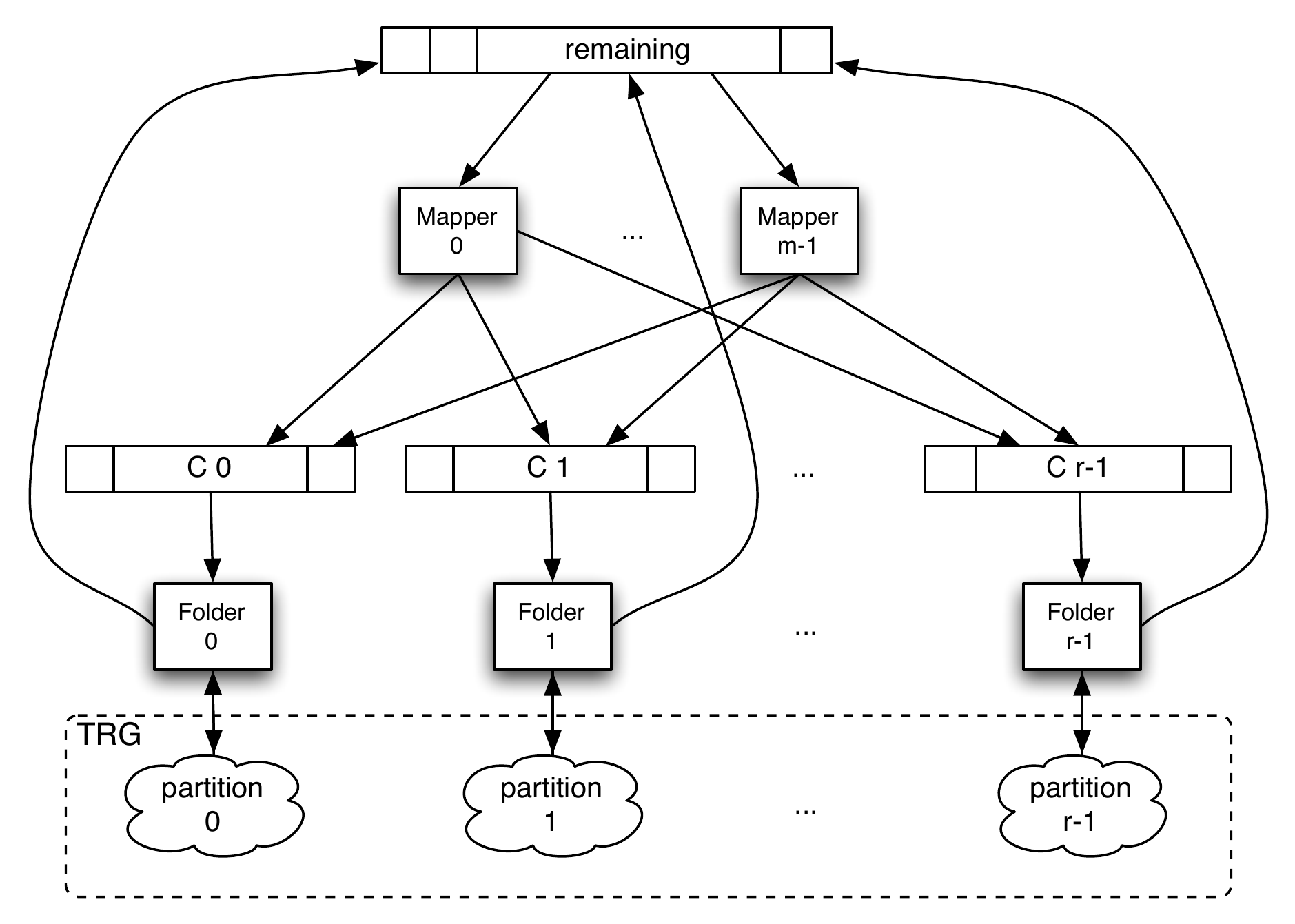}
\caption{\texttt{Mapper}s \&  \texttt{Folder}s model}
\label{fig:mt2}
\end{figure}

The second model specializes the \texttt{Worker}s in
\texttt{Mapper}s and \texttt{Folder}s (see Fig.~\ref{fig:mt2}). A
\texttt{Mapper} computational unit takes nodes from the expansion front, it maps them into their successors, and assigns the map outcome to the proper \texttt{Folder}s by means of the Hash function (\ref{eq:partitioning}) where $n$ is the number of \texttt{Folder}s; they in turn identify possible inclusion relationships, and build partitions of the whole final graph.

\vspace{.5cm}

It is worth noting that with respect to ordinary state-space
exploration techniques, both parallel models
incur in additional overheads due to extra communication and synchronization, greatly affecting
speed-up. The main overheads are due to the frequent locking of the
structure recording symbolic nodes (usually represented as hash
tables), and to the load imbalance deriving from asymmetric
computations performed by \texttt{Worker}s. 

\begin{figure}[htbf]
\centering
\includegraphics[width=1\columnwidth]{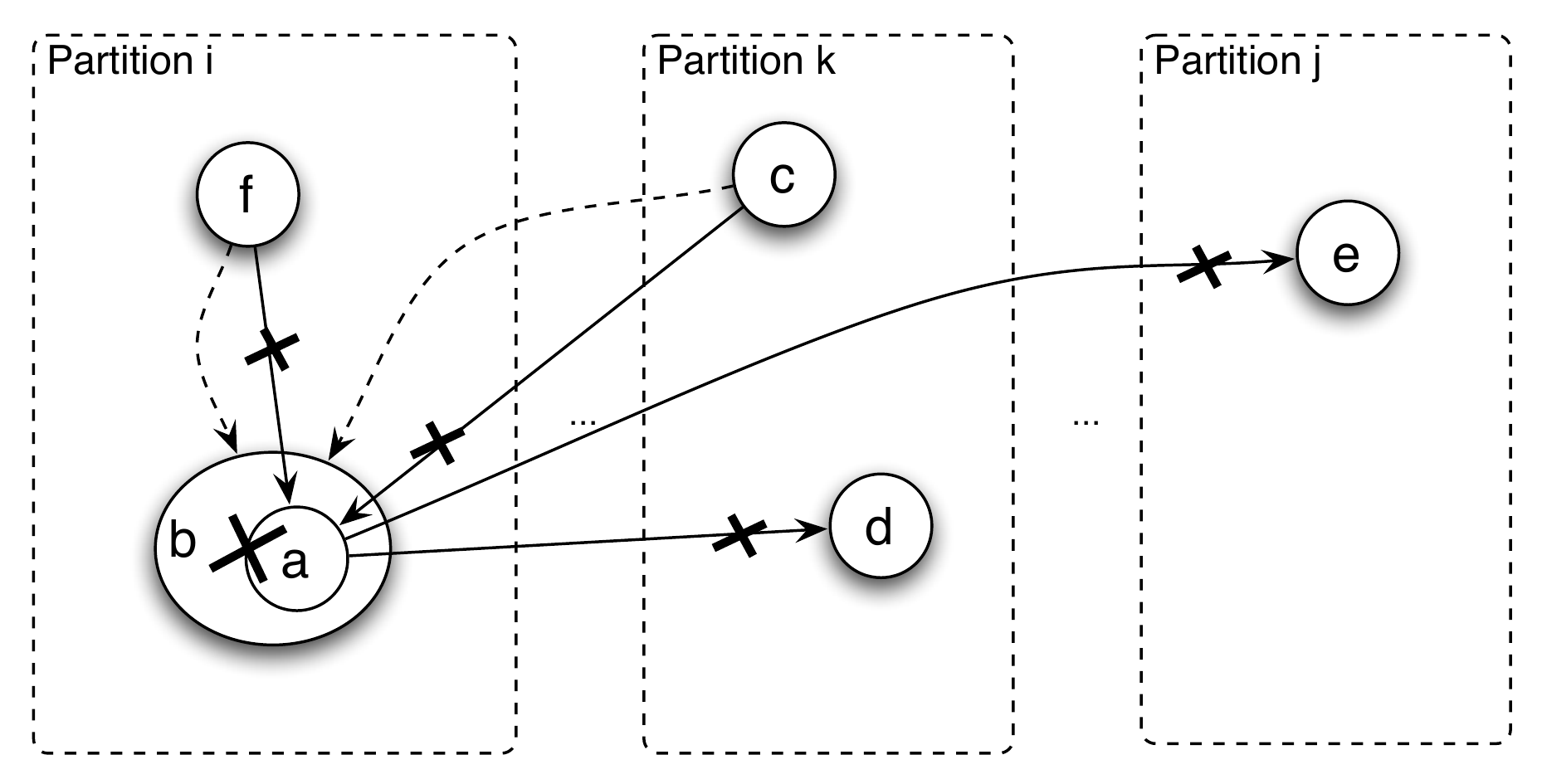}
\caption{Operations on edges during the \texttt{Fold} phase.}
\label{fig:edges_problem}
\end{figure}

Therefore, the conceptually global symbolic structure (the TRG) is partitioned among different computational units, according to the rule that each unit stores a set of nodes associated with the \emph{incoming} edges.
This choice makes the distributed management of the TRG easier: the only synchronization point is raised by the erasure (due to absorption) of nodes with outgoing edges. These information is not locally
present because outgoing edges are stored in the target nodes, which
are potentially belonging to other units. To minimize further the
communications between computational units, we perform a delayed removal
of pending edges (outgoing edges of removed nodes) at the end of the
global computation. For instance, the node $a$ represented in Fig. \ref{fig:edges_problem} is included in $b$. The redirection of the incoming edges ($f \rightarrow a$ and $c \rightarrow a$) can be performed locally because $a$ and $b$ belong to the same partition. The removal of outgoing edges ($a \rightarrow e$ and $a \rightarrow d$), instead, cannot be performed locally because $e$ and $d$ are not present in the partition $i$.

\section{Distributed implementations}
\label{sec:distr}

In order to scale to a large number of computational units we
considered different distributed architectures. In particular we used two
existing frameworks: JavaSpaces~\cite{Freeman99} and Hadoop
MapReduce~\cite{hadoop}. In this way we concentrated on the functional
aspects of our distributed application, while leaving to the frameworks the management of fault tolerance and low-level communication. While the \emph{JavaSpaces} implementation has been designed to run on local networks, 
\emph{MapReduce} has the possibility to be deployed ``in the cloud" in order to easily exploit a larger number of machines with better installed hardware.


\subsection{JavaSpaces Tool}
\label{sec:jspaces}
JavaSpaces technology is a high-level tool for building distributed
applications, and it can also be used as a coordination tool. It has
its roots in the Linda coordination language~\cite{Gelernter92}.
Departing from more traditional distributed models that rely on message passing
or RMI, the JavaSpaces model views a distributed application as a
collection of processes that use a persistent storage  (one or more
\emph{spaces}) to store objects and to communicate.

%

\begin{figure}[htbf]
\centering
\includegraphics[width=0.9\columnwidth]{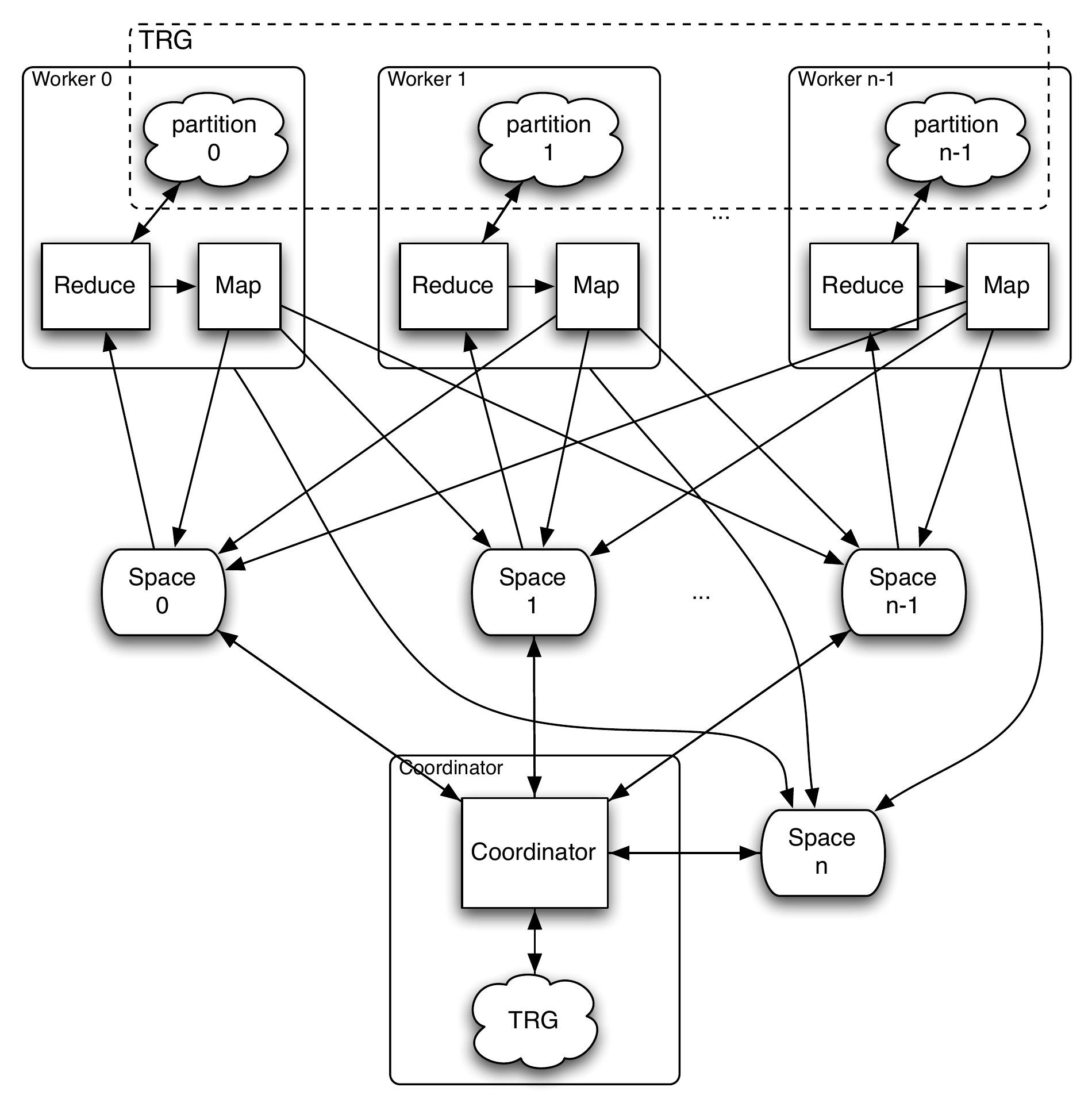}
\caption{Distributed JavaSpaces model}
\label{fig:jspacesmodel}
\end{figure}

By using this framework we have implemented the first parallel model
presented in Section~\ref{sec:first-model} (Fig.~\ref{fig:mt1}). We implemented each
\texttt{remaining} list as a space where \texttt{Worker} processes can
exchange states not yet examined. There is also one coordinator process
that initializes the computation by producing the initial state, then waits for the termination of each \texttt{Worker} in order to merge the computed partitions into the final TRG. The overall architecture is presented in Fig. \ref{fig:jspacesmodel}.

%
%
%
%

\subsection{Hybrid Iterative MapReduce}
\label{sec:HIMapred}
This is a distributed implementation of the second parallel model
presented in Section~\ref{sec:second-model} (Fig.~\ref{fig:mt2}). MapReduce is a well known programming model with associated implementations, for writing applications that rapidly process vast amounts of data in parallel on large clusters of computational cores. Users specify a \texttt{Map} function that processes a key/value pair to generate a set of intermediate key/value pairs, and a \texttt{Reduce} function that merges intermediate values associated to the same intermediate key.

\begin{equation} \label{eq:mapred}
\begin{split}
\texttt{Map}(k_{1},v_{1}) \rightarrow list(k_{2}, v_{2}) \\
\texttt{Reduce}(k_{2}, list(v_{2})) \rightarrow list(v_{2})
\end{split}
\end{equation}

In order to exploit this programming model we represent our data set as
pairs $\langle f(S),S\rangle$ , where $S$ is a node of the symbolic $TRG$
with associated incoming edges, $f(S)$ is the soft marking defined in (\ref{eq:partitioning}).

We actually used an extended version of the original MapReduce model introduced in \cite{Dean08}. With respect to such a model, MapReduce jobs are iterated until the  \emph{expansion front} becomes empty. This is called ``Iterative MapReduce'' \cite{Ekanayake10}. Each iteration maps all nodes in the expansion front, then it reduces the new nodes by identifying possible inclusion relationships. Note that the reduce phase requires all the $TRG$ nodes in order to identify each potential
inclusion relationship between them. For this reason, the input of each iteration is made up by a set of $new$ nodes  (the expansion front) and a set of $old$ nodes (the $TRG$ portion till now computed).

A \texttt{Map} takes a pair $\langle f(S),S\rangle$ as input. If it corresponds to an \emph{old}
node it is just passed to the reduce phase, without being processed.
Otherwise, the set $\{\langle f(S'),S'\rangle\}$ of the states directly 
reachable from $S$ is computed, and it is passed to the reduce phase
together with $\langle f(S),S\rangle$ itself.
After the map phase is concluded,
an intermediate \texttt{shuffle} phase brings together pairs with the same soft marking
($f(S)$) and it gives each group to a different \texttt{Reduce}. A
\texttt{Reduce} erases the values (states) that are shown to be included in some other and
it gives in output a set of values forming a partition of the $TRG$.

\begin{figure}[htbf]
\centering
\includegraphics[width=1\columnwidth]{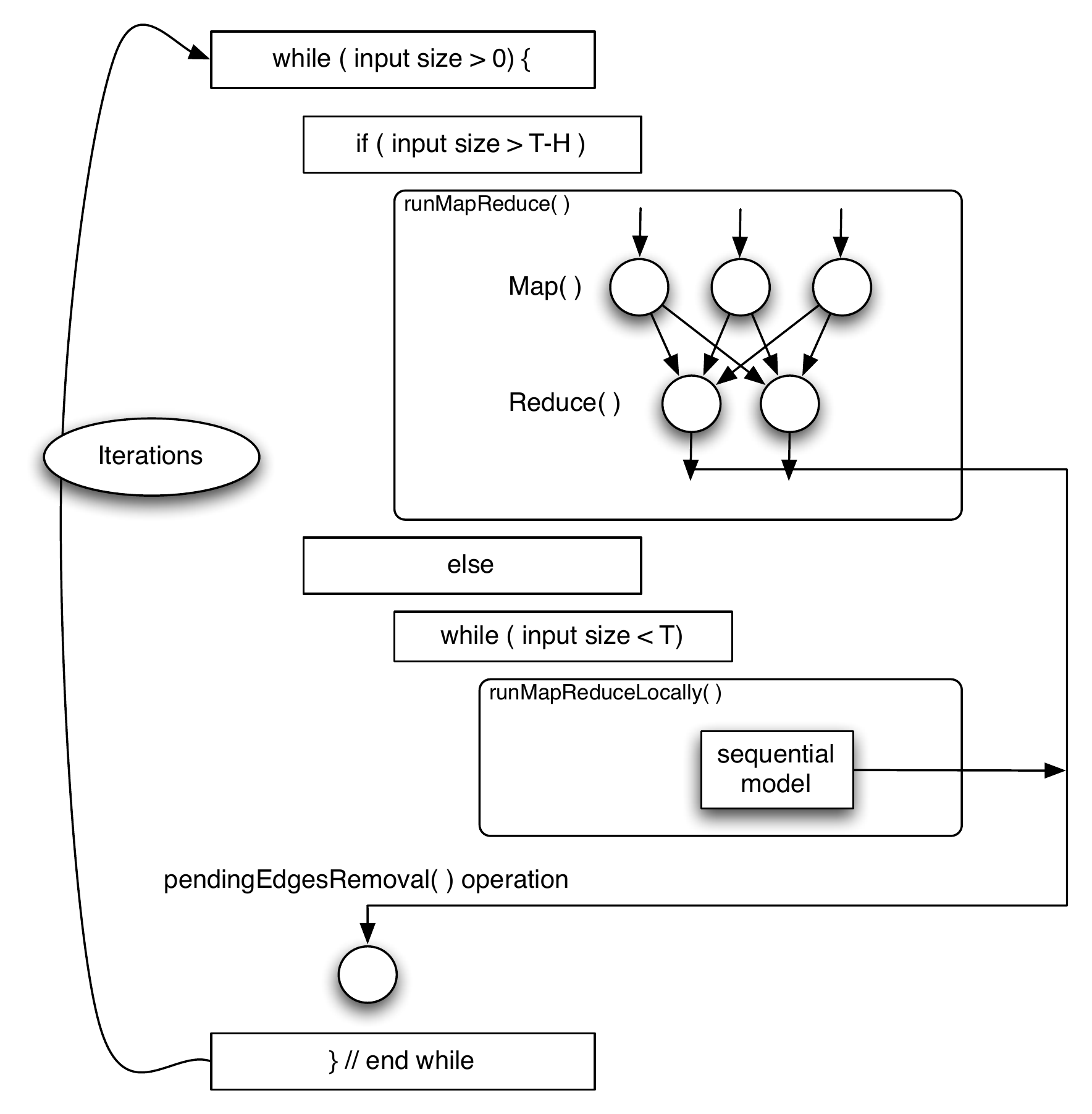}
\caption{Hybrid Iterative MapReduce model}
\label{fig:himapred}
\end{figure}

The original MapReduce model also permits one to define a \texttt{Combine}
function that performs a sort of local reduce on each \texttt{Map}'s
output, before the actual, distributed reduce phase. A
\texttt{Combine} runs on the same machine as the related \texttt{Map} and it
tries to partially aggregate intermediate data in order
to improve the overall system performance. In our application we have chosen
to discard this optimization because in TB nets context it is unlikely
that symbolic states generated by the same parent have the same marking~\cite{Bellettini11}.
Thus, a combine phase before the reduce phase could even affect the performance of our application. By the way,
using other formalisms this observation might be no more valid, and the \texttt{Combine} phase could be helpful.

Since the MapReduce model is not the best choice for elaborating a relatively small input, we introduced the possibility of changing the computational model, depending on the size of analyzed data set. Since the expansion front varies considerably during the $TRG$ building, it is convenient to use a sequential model on a single machine as long as it remains below a given threshold $T$. When the expansion front exceeds $T$, an Iterative MapReduce model on a large cluster of machines is employed. We call this approach (sketched in Fig. \ref{fig:himapred}) \emph{Hybrid Iterative MapReduce} (himapred).
A hysteresis ($H$) is also programmed, in order to react with some delay in front of possible swings of the expansion front within $T$.

\begin{figure}[htbf]
\centering
\includegraphics[width=0.7\columnwidth, angle=-90]{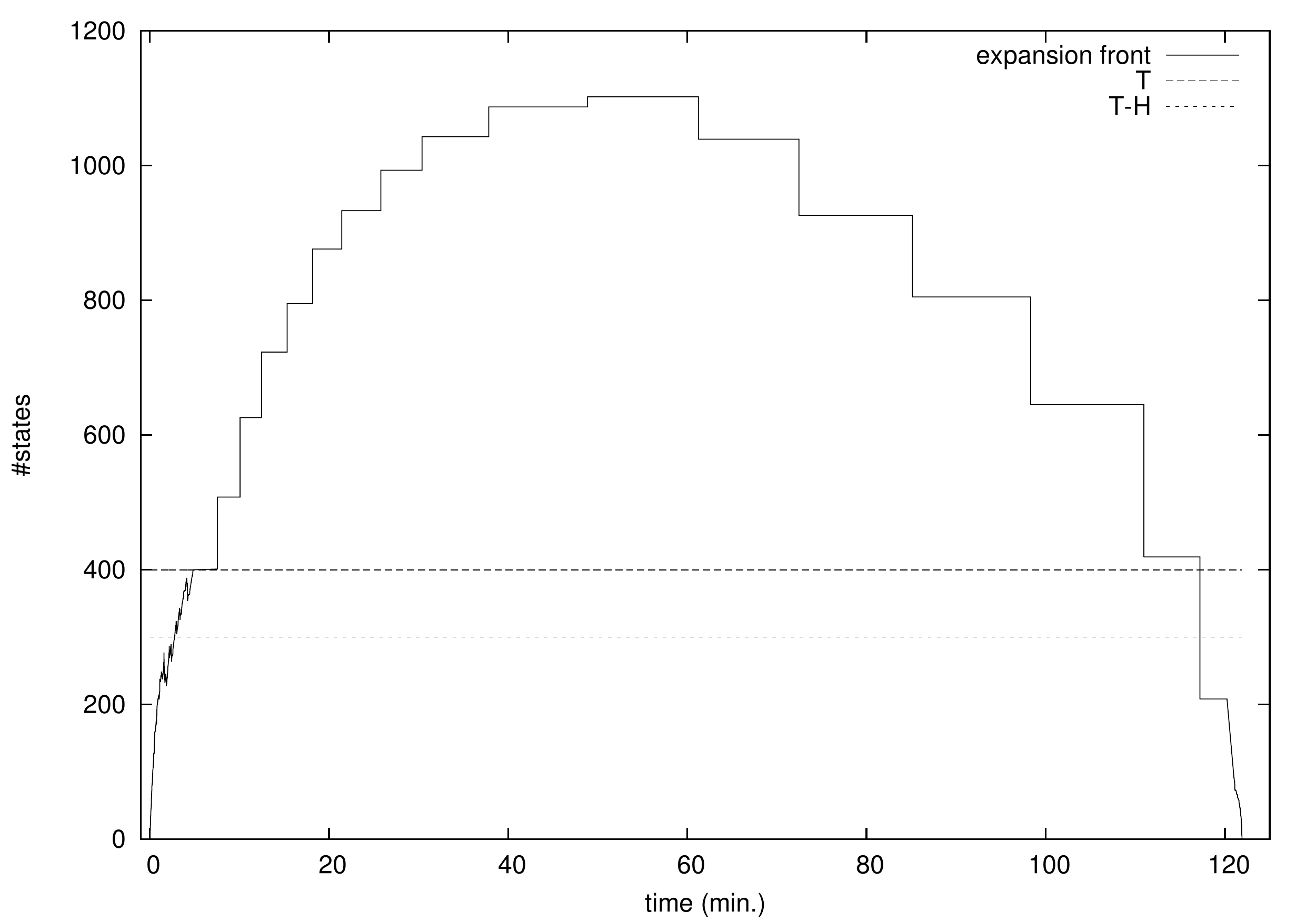}
\caption{Expansion front over time}
\label{fig:himapred_border}
\end{figure}

Fig.~\ref{fig:himapred_border} shows the expansion front of the Gas
Burner analysis over time. The trend line clearly shows how the execution
time of a single MapReduce iteration depends on the  $TRG$ size,
denoted $|TRG|$. Since a \texttt{Map} processes a single
sate, its execution time is independent from $|TRG|$ and in many cases it may be neglected.
Conversely, a \texttt{Reduce} works on a partition of
the $TRG$ (checking relationships between any pairs of nodes), thus its
complexity is $O(|TRG|^{2})$. The worst case occurs when all
nodes in the $TRG$ have the same feature $f(S)$: in that case a single
\texttt{Reduce} has to process the whole graph. Although the worst case is
very unlikely, a common situation is the presence of large clusters
of nodes that share the same key $f(S)$. This leads to a
computational load imbalance among the reducers often resulting in
a significant degradation of performances.

\section{Experiments}
\label{sec:exp}

The sequential builder produces a graph with 14563 nodes for the Gas
Burner example (versus 23635 symbolic states generated during
computation), and takes about 7.5 hours on a notebook with a 2.4Ghz
Intel Core 2 Duo processor and 4GB of RAM (the operating system is
Ubuntu 10.10 and the JVM is OpenJDK IcedTea6 1.9.5). In this paper we
adopt the Gas Burner example as a well known benchmark and we are not
interested in the properties of the system.

Testing activities on the \emph{JavaSpaces Tool} have been performed on a local network (33 computers over a 100Mb Ethernet LAN).
Preliminary experiments in this setting show that although performances are much
better than for the single-thread program (the execution time is reduced
by a factor $\sim3.75$), there is a major bottleneck preventing further improvements:
the state space partitioning among the \texttt{Worker}s set is not uniform. This means that some computation units are much more loaded than others, which remain idle for most of the time. 
In order to alleviate this problem, we conceived a different partitioning policy that allows for a higher degree of parallelism.
We used the function defined in (\ref{eq:partitioning}) with a different $f$, called \emph{discriminant soft marking}.
Let $dm$ be a function:

\begin{equation} \label{eq:disc_soft_marking_p}
\begin{split}
dm: P \rightarrow \mathbb{N}^{2}, ~dm(p) = \langle i,j \rangle
\end{split}
\end{equation}
where $p$ is a place
of the analyzed TB
net, $j$ is the number of \emph{anonymous} time-stamps in $p$, and $i$
is the number of \emph{other} time-stamps in $p$. The discriminant soft marking of $S$ is now defined as:

\begin{equation} \label{eq:disc_soft_marking}
\begin{split}
f(S) = \langle dm(p_{1}), ..., dm(p_{k}) \rangle \, \in  \mathbb{N}^{2k}
\end{split}
\end{equation}
This new definition comes from the observation that, even if two states have the same soft marking,
according to (\ref{eq:soft_marking}), they cannot be included into one another if the distribution
of \emph{anonymous} time-stamps in the corresponding markings is different.

\begin{figure}[htbf]
\centering
\includegraphics[width=0.7\columnwidth, angle=-90]{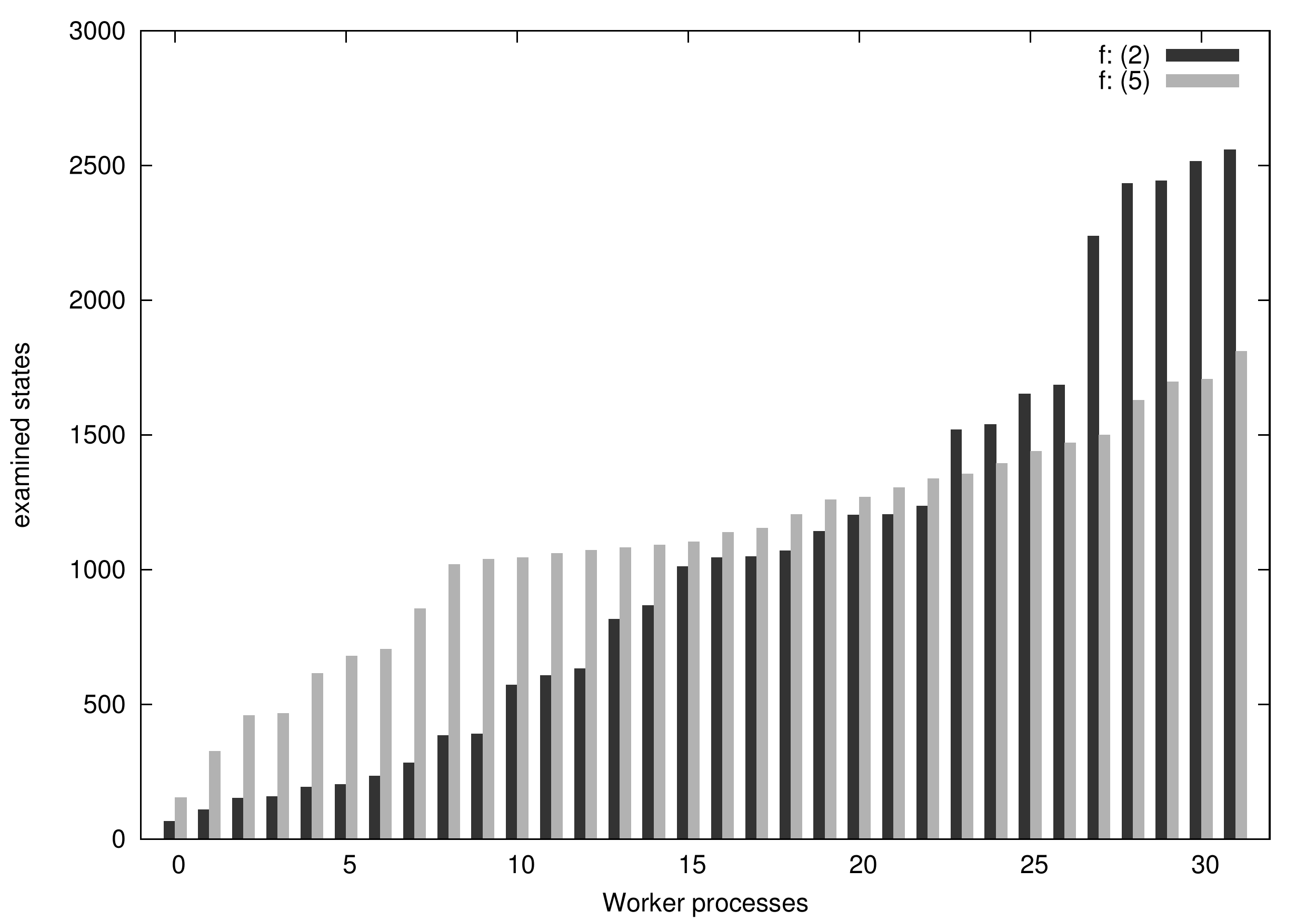}
\caption{State space partitioning among 32 \texttt{Worker}s}
\label{fig:partitioning}
\end{figure}

Fig.~\ref{fig:partitioning} shows the state space partitioning among 32 \texttt{Worker} processes using the two different partitioning policies. Table \ref{tab:ec2outcomes} reports the results of different experiments done within different settings.


\begin{table*}
\caption{Experiments Report}
\centering
{ \begin{tabular}{ | c | c | c | c | c | c | c | c | }
 \hline
  architecture & $\#$ compute units & tool version & compute model & T & H & f & exec. time\\
  \hline
  \hline
  2.4Ghz Intel Core 2 Duo, 2GB RAM & 1 machine & sequential & local (single machine) & - & - & (\ref{eq:soft_marking}) & $\sim$7.5 hrs\\
  \hline
  3Ghz Intel Pentium 4, 2GB RAM & 33 machines & JavaSpaces & local (distributed) & - & - & (\ref{eq:soft_marking}) & 1h55m40s\\
  \hline
  3Ghz Intel Pentium 4, 2GB RAM & 33 machines & JavaSpaces & local (distributed) & - & - & (\ref{eq:disc_soft_marking}) & 1h2m0s\\
  \hline
  m2.2xlarge  \cite{amazonElasticMapReduce} & 39 ec2 units & himapred & cloud & 200 & 50 & (\ref{eq:soft_marking}) & 1h35m33s\\
   \hline
   m1.xlarge  \cite{amazonElasticMapReduce} & 104 ec2 units & himapred & cloud & 200 & 50 & (\ref{eq:soft_marking}) & 1h43m19s \\
   \hline
   m2.2xlarge  \cite{amazonElasticMapReduce} & 104 ec2 units & himapred & cloud & 200 & 50 & (\ref{eq:soft_marking}) & 1h0m0s \\
   \hline
   m2.2xlarge  \cite{amazonElasticMapReduce} & 104 ec2 units & himapred & cloud & 400 & 100 & (\ref{eq:disc_soft_marking}) & 46m8s \\
   \hline
   m2.2xlarge  \cite{amazonElasticMapReduce} & 104 ec2 units & himapred & cloud & 200 & 50 & (\ref{eq:disc_soft_marking}) & 39m33s \\
   \hline
\end{tabular}}
\label{tab:ec2outcomes}
\end{table*}

The last  \emph{MapReduce Tool} has been deployed ``in the cloud'' by means of the Amazon Elastic MapReduce web service \cite{amazonElasticMapReduce} that employs the Amazon Elastic Compute Cloud (EC2) infrastructure. Table ~\ref{tab:ec2outcomes} summarizes the outcomes of the Gas Burner analysis carried out using different distributed frameworks with varying configurations.
The results point out the different factors that contribute to improve the performances of our distributed applications:
the number of computation units, the cluster dimension, the hardware of each cluster machine, and the partitioning policy.
In particular the latter one shows to be a key factor for the possibility of conveniently scaling up the available computation resources.

\section{Conclusion and future works}
\label{sec:conc}

We have presented and discussed some approaches based on exploitation of distributed/cloud computing frameworks to deal with the state-space explosion in real time system analysis .
These approaches have been experienced for timed (symbolic) reachability analysis of Time Basic (TB) Petri nets.
The proposed implementations extend the sequential builder of TB nets' time reachability graph.
Standing on a common basic computational schema (a sort of Map \& Fold), our approach is general enough to be used within different formalisms by specializing the $state$, the \texttt{Map}, the \texttt{Reduce}, and the $f$ concepts.
In particular, we have designed and implemented an extension of the MapReduce model,
called Hybrid Iterative MapReduce. 
The outcomes of tests performed on a benchmarking RT model
clearly show how distributed (especially cloud) implementations can be conveniently
used to increase the performances of the sequential builder. 
We plan to extend our research by trying to further refine the partitioning function and 
studying ways for integrating dynamic load balancing models not only into the JavaSpaces implementations
but also into iterative MapReduce based computational frameworks,
in order to cope with the main performance bottleneck.

Examples and binaries of the tools described in this paper can be
found at: \url{http://camilli.dico.unimi.it/graphgen}
with associated ``how to install'' notes.


\bibliographystyle{plain}
\bibliography{biblio}


\end{document}